\documentclass{kapproc} 
\usepackage{epsfig}
\kluwerbib  

\newcommand\beq{\begin{equation}}
\newcommand\eeq{\end{equation}}
\newcommand\bea{\begin{eqnarray}}
\newcommand\eea{\end{eqnarray}}

\begin{document}
\articletitle{Passive fields and \mbox{particles }\\
in chaotic flows
}

\author{Bruno Eckhardt, Erwan Hasco\"et, Wolfgang Braun}
\affil{Fachbereich Physik\\
Philipps Universit\"at Marburg\\
35032 Marburg\\
Germany\\
}
\email{bruno.eckhardt@physik.uni-marburg.de}

\begin{abstract}
Two examples for the interplay between chaotic dynamics and 
stochastic forces within hydrodynamical systems are considered.
The first case concerns the relaxation to equilibrium of
a concentration field subject to both chaotic advection and
molecular diffusion. The concentration field develops filamentary
structures and the decay rate depends non-monotonically on the
diffusion strength. The second example concerns polymers, modelled
as particles with an internal degree of freedom, in a chaotic
flow. The length distribution of the polymers turns out to
follow a power law with an exponent that depends
on the difference between Lyapunov exponent and internal
relaxation rate. 
\end{abstract}


\section*{Introduction}
Fluid dynamics provides many prime examples for nonlinear dynamical
systems. The origin of the nonlinearity is the advection of 
particles and fields by the underlying velocity field and takes,
with ${\bf x}_P(t)$ the position of a particle, the form
\beq
\dot {\bf x}_P(t) = {\bf u}({\bf x}_P(t),t)\,.
\eeq
We assume here and in the following that the particles are
neutrally buoyant so that they follow the instantaneous
velocity field (for other particles, see \cite{Maxey83}).
Typically, the fields are not constant or linear,
so that (1) is a system of nonlinear equations and the
dynamics of the particles becomes chaotic
(\cite{Aref84,Ottino89,SF91,CS94,Aref02}).
Stochastic elements enter if in addition to the deterministic parts also
random forces, such as molecular diffusion or Brownian motion, have
to be taken into account. 
Further complications arise when the 
particles can react back onto the flow (e.g. by a density contrast
or by modifying viscosity). They then become dynamically active. 
In the situations considered below no such feedback is allowed and
the particles remain dynamically passive.

Here we consider two examples of the interplay between nonlinear
advection and stochastic forcing: the relaxation of scalar fields in
experiments with 2-d flows (\cite{Gollub99,Gollub97,Tabeling97}) and
the stretching of polymers in chaotic flows.  In all cases the flow
fields have a simple spatial structure and at most a periodic time
dependence, i.e. we have deterministic chaos. If these deterministic 
flow fields are replaced by temporally fluctuating ones with prescribed
spatial correlations techniques from stochastic analysis can be
used to characterize the particle dynamics further 
(\cite{Falkovich02,Klyatskin96,Cranston02}). However, both
deterministic and random flows show similar behaviour.

In the first example we focus on the processes that contribute
to the relaxation of a non-uniform concentration,
e.g. a blob of dye, to a uniformly spread out concentration in a chaotic
flow. By its chaotic nature the flow by itself will stir the 
blob into an object with filaments that grow longer and finer and
eventually cover space. Diffusion alone will also
result in a uniform spreading in space, albeit on a different time
scale. When combined, these processes do not simply add up but
give rise to non-trivial modulations in the relaxation rate.

In the second example we focus on the effects of the velocity gradients on 
particles with internal degrees of freedom, specifically long flexible 
polymers. The phenomenon of turbulent drag reduction
(\cite{Lumley69,Virk75}) is connected
with the uncoiling of the polymers in the flow. Once uncoiled, the polymers
are no longer passive but begin to change the velocity field, so that
a reduced drag is observed. The passive dynamics of polymers in
a prescribed flow field then is a first step towards a description 
of the polymer dynamics in a turbulent flow. It is also a nice
example of how the presence of both additive and multiplicative
fluctuating forces can give rise to a power law probability
distribution (\cite{Balkovsky00,Balkovsky01,Chertkov00,Sornette97}).

\section*{Relaxation in chaotic advection}
Since the paper by Aref (\cite{Aref84}), in which the term chaotic
advection was coined, many theoretical and experimental studies have
addressed various aspects of the phenomenon
(\cite{Ottino89,SF91,CS94,Gollub97,Gollub99}). 
However, in order to arrive at mixing on a molecular level,
diffusion has to be added in. Already in
(\cite{Eckart48}) the differences between stirring and 
mixing are emphasized: during stirring the fluid elements are 
stretched and folded, but they always keep their identity.  Mixing
requires molecular diffusion to cause a diffusive exchange of
markers between different fluid elements and to thus blur
distinguishability of the origin of the fluid elements. It is obvious
that stirring enhances mixing, but there will be mixing even
without stirring. The persistent patterns proposed in
\mbox{(\cite{Pierrehumbert94})} and observed experimentally in (\cite{Gollub99})
and the studies by (\cite{Vassilicos02}) highlight some of the 
effects that arise when chaos and diffusion interact to turn
stirring into mixing.

For fast processes the effects of diffusion can be neglected.  Typical
times for the stirring by the flow are set by $L/U_0$, where $U_0$ and
$L$ are characteristic velocity and length scales,
respectively. Diffusive spreading happens on a time scale $L^2/D$,
where $D$ is the diffusion constant, and is usually much slower. The
experiments by Taylor on reversibility at low Reynolds number 
(\cite{Taylor60,Homsy00}) also
indicate this vast separation of time scales: the flow is reversed
before diffusion has time to destroy the memory of the initial
conditions  (a further discussion of such echo phenomena in
chaotic flows is given in (\cite{Eckhardt03})).

\begin{figure}
\begin{minipage}{0.68\textwidth}
\epsfxsize\textwidth
\epsfbox{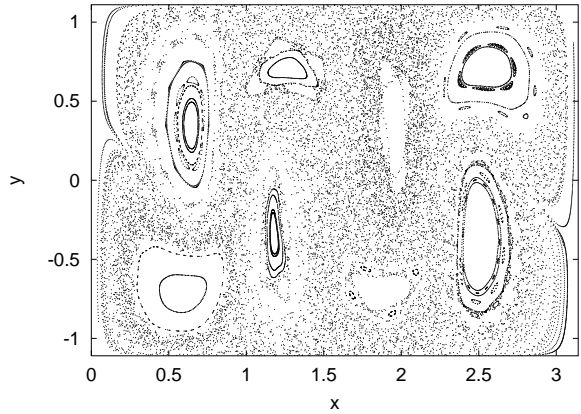}
\end{minipage} \hfill
\begin{minipage}{0.29\textwidth}
\caption[]{Stroboscopic map for the flow (\ref{modes}). The map was 
obtained by iterating 100 uniformly distributed initial conditions
for 200 periods each and plotting all intermediate points.
The frequency of the forcing is $\omega=20$ and its amplitude $f_0=300$.}
\label{strobmap}
\end{minipage}
\end{figure}

\begin{figure}[b]
\begin{minipage}{0.68\textwidth}
\epsfxsize\textwidth
\epsfbox{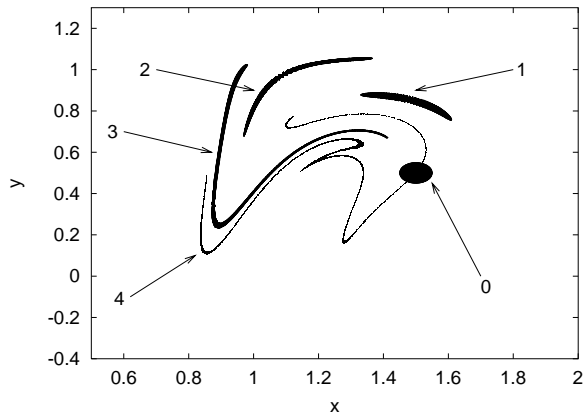}
\end{minipage} \hfill
\begin{minipage}{0.29\textwidth}
\caption[]{Evolution of a small blob of ten thousand particles
after a few periods. Labels indicate multiples of a
period: $t=nT$ with $n=0$ (initial configuration) and $n=1$, $2$, $3$ and 
$4$, after which time one tendril of the stretched out blob overlaps 
with the initial configuration. The parameters are frequency
$\omega=40$ and the forcing amplitude $f_0=1500$.}
\end{minipage}
\label{sequence}
\end{figure}

However, this estimate has to be reconsidered when the dynamics 
builds up finer and finer structures and increasing gradients 
so that diffusion can no longer be neglected (cf.
\cite{Pierrehumbert94,Vassilicos02}). 
Consider particle spreading in a 2-d flow designed to 
describe the experiments of (\cite{Gollub97,Gollub99}). 
The stream function is
(\cite{Eckhardt02a})
\bea
\psi(x,y,t) &=& \frac{f_0}{3\sqrt{9+\omega^2}} \sin(\omega t - \delta_1) 
	 \sin x\, \cos \sqrt{2} y \nonumber\\
&\ & + \frac{5 f_0}{27\sqrt{729+\omega^2}} \sin(\omega t - \delta_2) 
	 \sin 5 x\, \cos \sqrt{2} y \,.
\label{modes}
\eea
where $f_0$ is the amplitude of the periodic forcing which sets the
velocity scale. The two modes respond to the driving with the phase
delays $\delta_1=\mbox{atan}(\omega/3)$ and
$\delta_2=\mbox{atan}(\omega/27)$, respectively, the difference in
$\delta_1$ and $\delta_2$ being a prerequisite for chaotic motion.
Fig.~\ref{strobmap} shows a 
Poincar\'e surface of section and Fig.~\ref{sequence} a sequence
of stroboscopic plots for many particles started
in a small area. The initial distribution is stretched
out along the unstable manifolds nearby and contracts along 
the stable ones. It soon folds back just as the manifolds do. 
After a few periods the particles are more or less uniformly
distributed in the chaotic region. 

The presence of islands in the flow adds a slow component
due to the trapping near islands.  The relaxation becomes
even slower when diffusion is added since 
stochastic perturbations can bring particles into islands
from which they can escape with difficulty only. Such localization
in islands is reflected in eigenvalues and eigenmodes
(O. Popovich and A. Pikovsky, in preparation).

But even if there are no islands diffusion can have
a nontrivial influence on the decay rate. In order to 
be able to vary the diffusion constant over a wide range and still
be numerically reliable we illustrate this possibility for 1-d maps 
on a ring, e.g.
\beq
f(x) = \left\{\matrix{3x & 0<x<1/3 \cr
2-3x & 1/3<x<2/3 \cr
-2+3x & 2/3 < x < 1}\right. \,.
\label{map}
\eeq
The map is periodically continued and noise is added to 
simulate molecular diffusion. The calculation proceeds by 
taking the Frobenius-Perron equation for the first
iteration step, followed by a Gaussian smearing.
The eigenvalues can be calculated by numerical
diagonalization in either position space or
Fourier space (Hasco\"et and Eckhardt, to be published).

\begin{figure}
\begin{minipage}{0.65\textwidth}
\epsfxsize\textwidth
\epsfbox{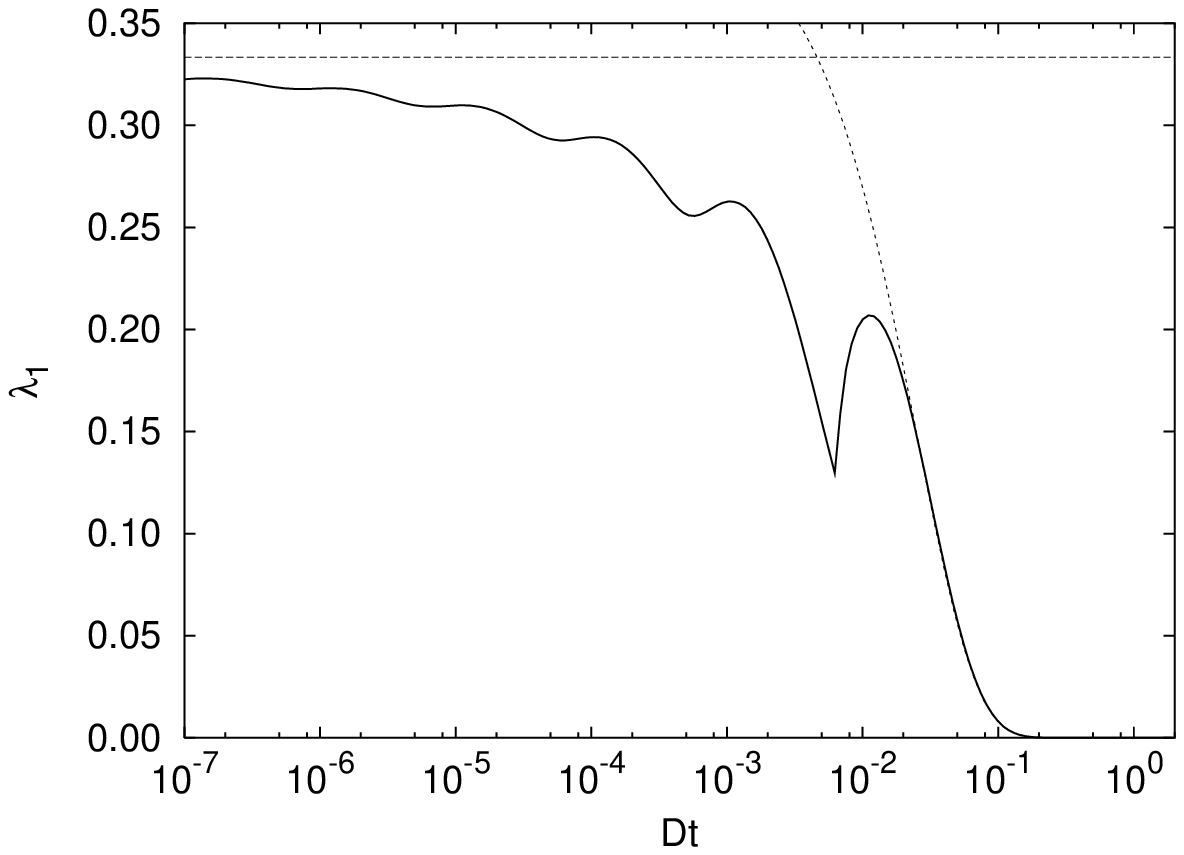}
\end{minipage} \hfill
\begin{minipage}{0.3\textwidth}
\narrowcaption{Variation of leading eigenvalue with 
diffusion for the map (\ref{map}). The calculated eigenvalue
is indicated by the thick line. For extremely
small diffusion it approaches the eigenvalue of the 
deterministic map (dashed horizontal line), and for large diffusion
the slowest one of the unperturbed diffusion equation
(dotted line). }
\label{spectrum}
\end{minipage}
\end{figure}

The spectrum of the unperturbed Frobenius Perron
equation can be calculated e.g. from periodic orbit
theory. The eigenvalue next to 1 comes out to be
1/3. Similarly, if diffusion is large enough, 
the deterministic part is a small perturbation and
we expect eigenvalues that depend on diffusion like
$\exp(-4\pi^2Dt)$, where $t$ is the time interval over
which diffusion acts in-between iterations of the map. 
The leading eigenvalue of the spectrum is shown
in Fig.~\ref{spectrum}.  Instead of a simple cross over
from fast chaotic relaxation for small diffusion to 
diffusional decay for large $D$ we see non-monotonic
variations with intriguing oscillations.

Similar non-monotonic behaviour has been observed in diffusion in
chains of tent maps when the height of the tip of the tent was
modified (\cite{Klages02,Klages99,Klages97,Klages95}). There, the
non-monotonic decay could be traced back to the non-monotonic changes
in the phase space structures, the homoclinic and heteroclinic tangles
and the changes in the efficiency that particles could be
exchanged with between different sets of the Markov partition. We expect a
similar mechanism here: the decay within the classical dynamics is
connected with the efficient stretching and folding. Diffusion can
help by mixing faster within cells, but it can also slow down 
relaxation by transporting fluid into regions with less efficient mixing.

\section*{Polymer stretching in chaotic flows}
The second example we consider is the 
dynamics of long flexible polymers (\cite{Lumley69,Virk75}). 
In fluids at rest these polymers are coiled up into
tiny spheres. However, despite their small size, they
can interact with gradients of the flow field
in turbulent and also in chaotic flows. 
When the strains are stronger
than their entropic relaxation forces they can be stretched out
to form elongated structures or even thin threads.
Elongated polymers influence the viscosity and change
the dynamics on the smallest scales. In the case
of turbulent drag reduction these microscopic
modifications add up to large scale changes in the 
velocity field that result in noticeable reductions
in drag (\cite{Lumley69,Virk75}).
As a part of a program to understand the effects 
of polymers on the flow we need to understand
the size distribution of the polymers.

In the simplest, and for our purposes sufficient model
for the polymer, the large number of monomers and 
flexible elements are replaced by their inertial
tensor, which can be written as a dyad formed 
by a vector ${\bf R}$, a kind of end-to-end 
distance vector (\cite{Bird87}). This vector is driven by thermal
fluctuations $\xi$, modelled as a 3-D Gaussian process. 
The entropic preference of the molecule
for a coiled state results in an entropic restoring
force, which contributes a relaxation time 
$\lambda$ towards the equilibrium configuration.
The small size of the polymers allows to linearize
the flow field, so that only local gradients enter.
We thus have a polymer transported along a Lagrangian 
path ${\bf x}_P(t)$, with an internal dynamics
described by
\beq
  \dot R_i=\left(\frac{\partial u_i}{\partial x_j}(t)-\frac 1{2\lambda}\right)
           R_j+\xi_i\,.
  \label{eq_model}
\eeq
From this vector one can form the 
configuration tensor, $c_{ij}=\langle R_i R_j \rangle$,
where the averages are over thermal noise.
The quantity used to measure extension of the polymers
is $\mbox{tr\,} c$. Scales can be fixed so that
in equilibrium $c_{ij}=\delta_{ij}$ and $\mbox{tr\,} c=3$.  

\begin{figure}
\epsfxsize0.49\textwidth
\epsfbox{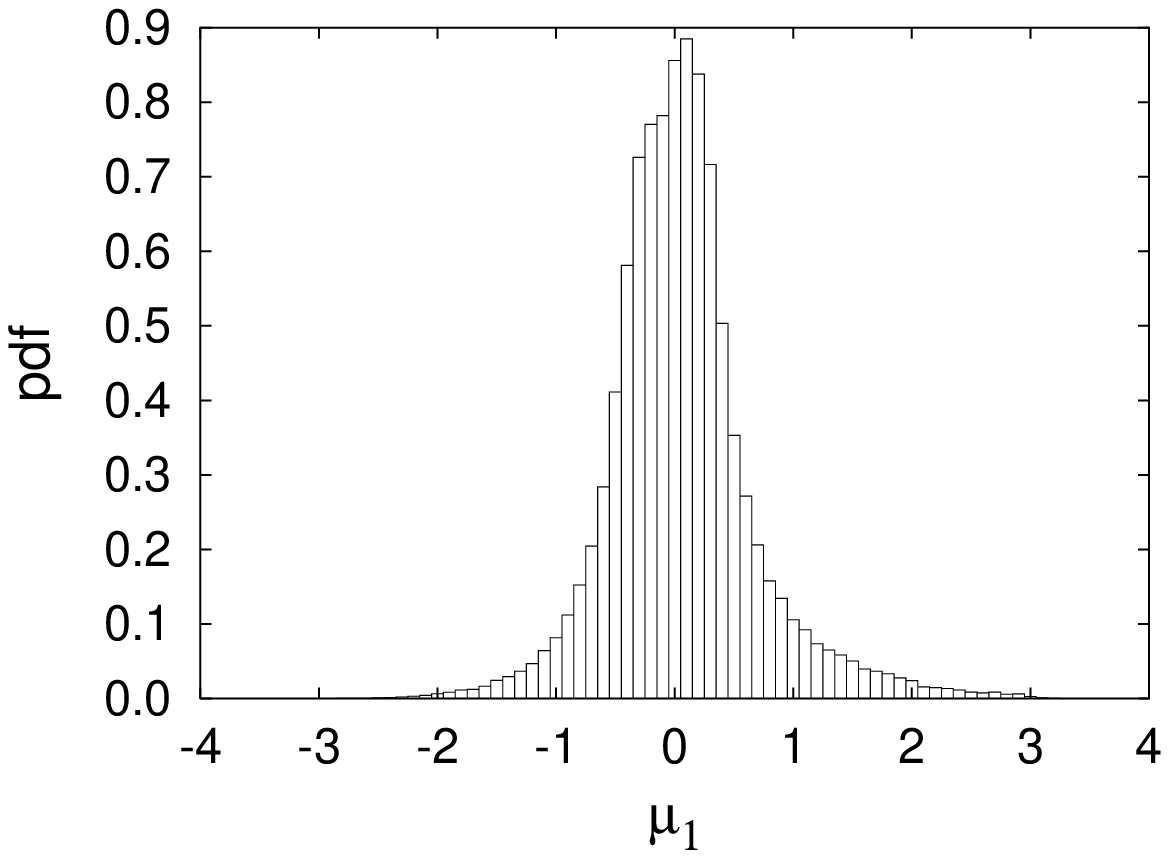}
\epsfxsize0.49\textwidth
\epsfbox{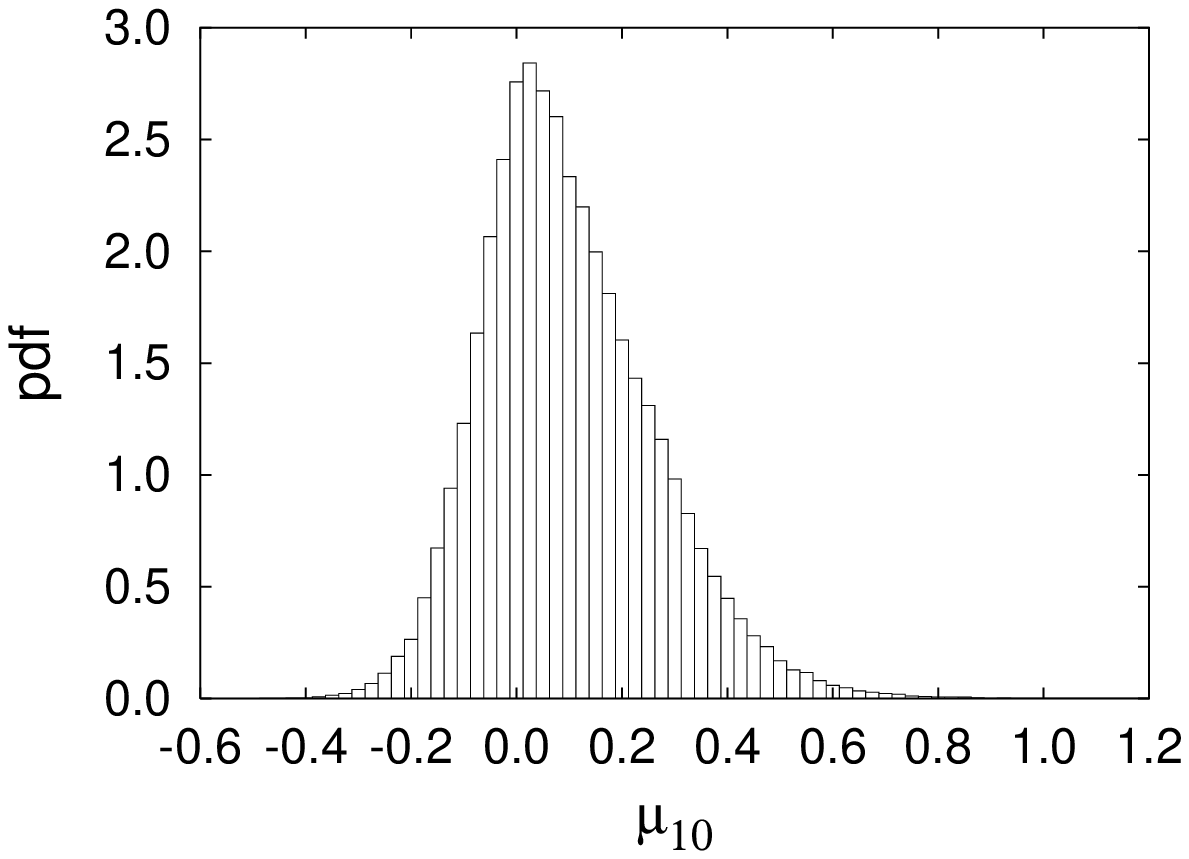}
\caption[]{Finite time Lyapunov exponents for two different time
intervals $T$ and the velocity field (\ref{eq_ABC}).  The 
mean in both cases is $\bar\mu=0.11$, and the empirical
variance decreases from 0.373 for $T=1$ to 0.026
for $T=10$.}
\label{finitemu}
\end{figure}

Since the polymers are advected by the flow
the local gradients vary. Asymptotically in the infinite
time average they approach the Lyapunov exponent of the
trajectory. In order to capture some of the variations
of the local stretching rate over times compatible with the
internal relaxation time of the polymers, we consider
distributions of finite time Lyapunov exponents,
\beq
 \mu_T = {1\over T} \log{{\|\delta\mathbf x(t_i+T)\|}\over
 {\|\delta\mathbf x(t_i)\|}}\,.
\eeq
They measure the growth rate of a separation 
$\delta \mathbf x$ over a time interval $T$.
The equation for $\delta\mathbf x$ is the variational equation
\beq
  \delta\dot x_i={\partial u_i\over \partial x_j}(t) \delta x_j\,.
\eeq
As the time $T$ increases the distribution narrows around the 
infinite time Lyapunov exponent (Fig.~\ref{finitemu}).

The length distribution of the polymers results
from a balance between the stretching by the gradients
of the velocity field, as measured by the
Lyapunov exponent $\mu$, and the entropic relaxation.
If the gradients and thus the Lyapunov exponents were
constant, the thermal fluctuations would give rise
to a Gaussian distribution. However, in the presence
of fluctuations in the Lyapunov exponent a power law
distribution of the size of the polymers follows
(\cite{Balkovsky00,Balkovsky01,Chertkov00}).
A corresponding result in a more general context
of multiplicative stochastic processes with a reflecting
barrier is given in (\cite{Sornette97}).
In fact, the noise term in (\ref{eq_model}) 
can be replaced by a reflecting barrier without
changing the tails of the polymer length distribution.
Evidently, in a power law distribution large
extensions of the polymer are much more likely than
in a Gaussian distribution and thus this higher
probability results in a more effective and efficient
back reaction of the polymers on the flow.

Since it makes no difference for the polymer dynamics if
the stretching results from a spatially and temporally
fluctuating turbulent flow or from a stationary flow 
with chaotic trajectories, we can also analyze the dynamics 
of polymers in a 3-d incompressible flow with chaotic
trajectories, i.e.
\beq
\mathbf{u}=\left(\matrix{(A\sin\pi y+B\sin\pi z)\sin\pi x\cr
                          A\cos\pi x\cos\pi y+C\cos\pi z \cr
                          B\cos\pi x\cos\pi z+C\sin\pi y } \right)
  \label{eq_ABC}
\eeq

\begin{figure}
\begin{minipage}{0.65\textwidth}
\epsfxsize\textwidth
\epsfbox{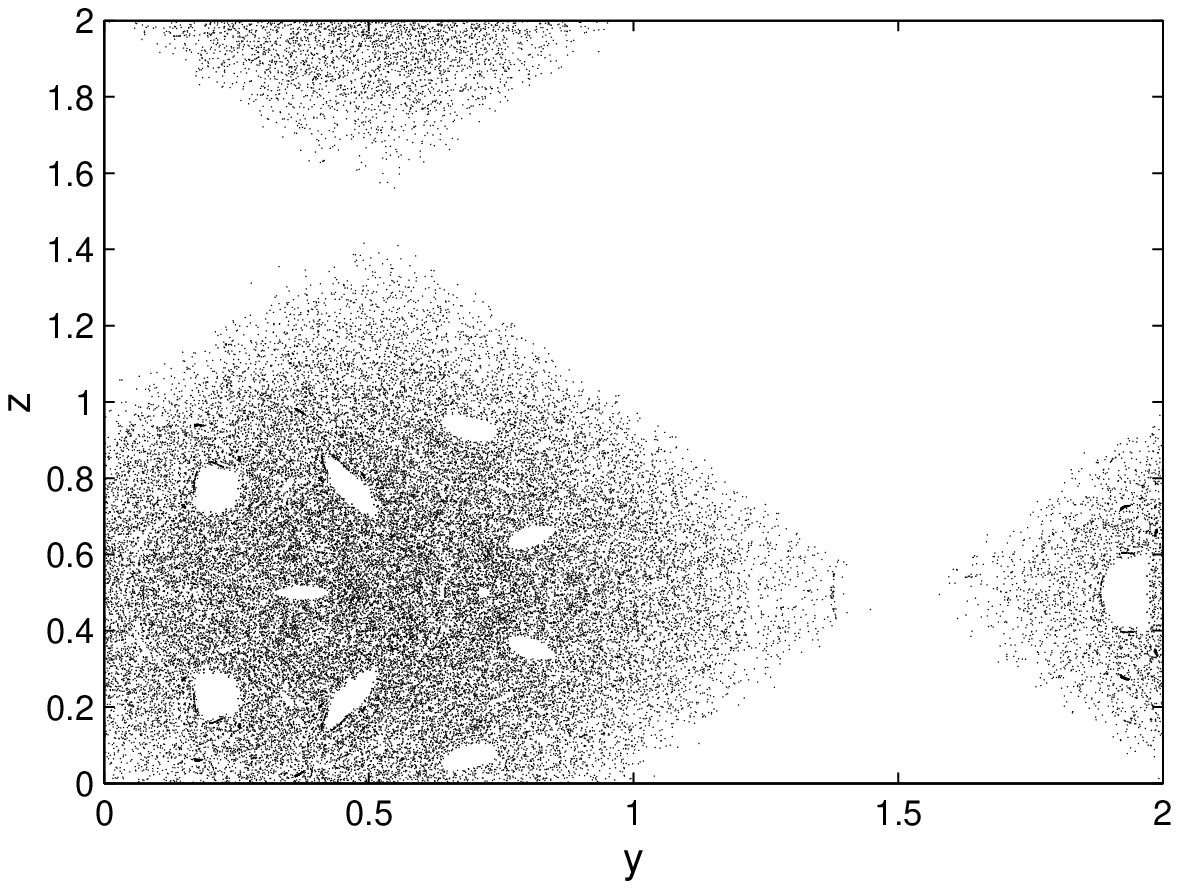}
\end{minipage} \hfill
\begin{minipage}{0.3\textwidth}
\narrowcaption{Poincar\'e surface of section for 
flow (\ref{eq_ABC}) with parameters $A=B=1$, $C=0.3$ at $x=0.5$. 
The non-uniform distibution of points is connected with a non-uniform
invariant density. The holes indicate elliptic islands.}
\label{abc_section}
\end{minipage}
\end{figure}

This flow is similar to the ABC flows (\cite{Dombre86}) but
satisfies free slip boundary conditions in the planes
$z=0$ and $z=1$. The particle trajectories are
trapped between the surfaces, as they would be in a planar
shear flow. 
Fig.~\ref{abc_section} shows a Poincar\'e surface of section
of a chaotic trajectory in this flow.
The distribution of finite time Lyapunov exponents
(Fig.~\ref{finitemu}) indicates a mean Lyapunov exponent 
of about $0.11$.

\begin{figure}
\epsfxsize0.49\textwidth
\epsfbox{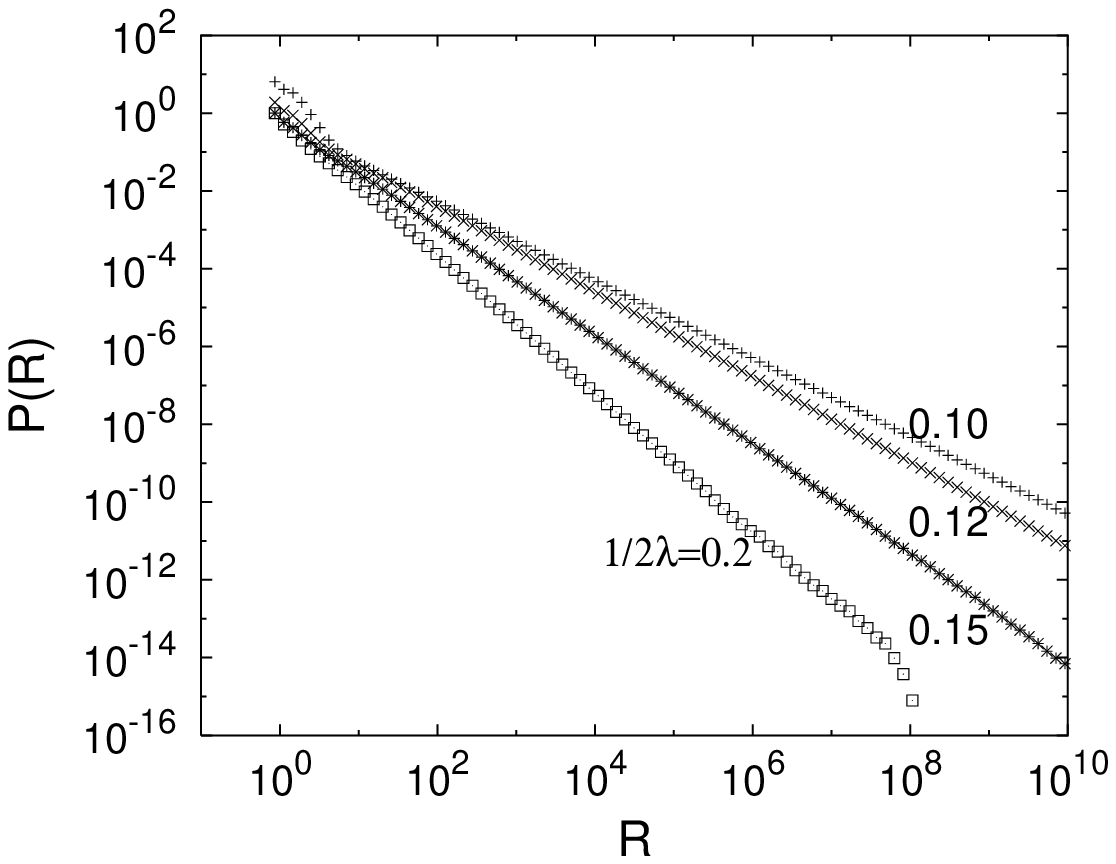}
\epsfxsize0.49\textwidth
\epsfbox{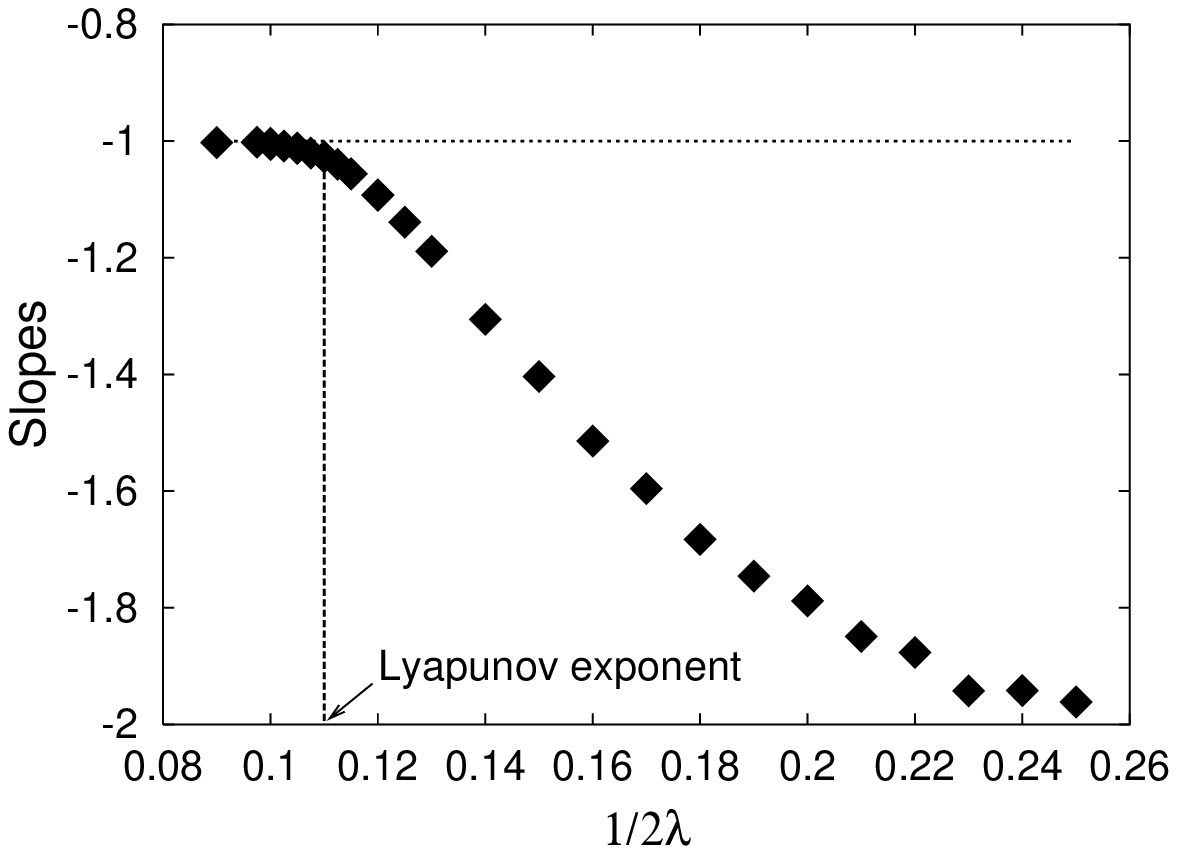}
\sidebyside
{\caption{Polymer length distribution (not normalized)
for different values of the polymeric relaxation time $\lambda$.
}\label{distribution}}
{\caption{Exponent in the distribution of $tr\,c$ vs. the inverse
of the relaxation time $\lambda$.}
\label{exponent}}
\end{figure}

As a result of the interplay between the exponential
stretching and its fluctuations one ends up with 
a power law distribution (Fig.~\ref{distribution})
with an exponent that depends on the relaxation rate $\lambda$. 
The distribution loses its normalizability (exponent
-1 or larger) when $2 \mu \lambda=1$. Numerical
calculations for the polymer length support this
(Fig.~\ref{exponent}).

The power law distribution has also been identified
in turbulent flows (\cite{Eckhardt02}). The consequences 
for the velocity field when the backreaction of the polymers to
the flow is allowed for are under investigation.

\section*{Acknowledgements}
We would like to thank 
O. Popovych and A. Pikovsky for stimulating
exchanges on relaxation in advection diffusion equations.
This work was supported in part by the EU within 
HPRN-CT-2000-00162 (Non-Ideal Turbulence) and by the 
Deutsche Forschungsgemeinschaft.

\bibliographystyle{apalike}

\begin{thebibliography}{}

\bibitem[Acrivos et~al., 1991]{SF91}
Acrivos, A., Aref, H., and Ottino, J.~M., editors (1991).
\newblock {\em Symposium on Fluid Mechanics of Stirring and Mixing}, volume~5
  of {\em Phys. Fluids A}.

\bibitem[Aref, 1984]{Aref84}
Aref, H. (1984).
\newblock Stirring by chaotic advection.
\newblock {\em J. Fluid. Mech.}, 143:1.

\bibitem[Aref, 1994]{CS94}
Aref, H., editor (1994).
\newblock {\em Chaos Applied to Fluid Mixing}, volume~4 of {\em Chaos, Solitons
  and Fractals}.

\bibitem[Aref, 2002]{Aref02}
Aref, H. (2002).
\newblock The development of chaotic advection.
\newblock {\em Phys. Fluids}, 14:1315.

\bibitem[Balkovsky et~al., 2000]{Balkovsky00}
Balkovsky, E., Fouxon, A., and Lebedev, V. (2000).
\newblock Turbulent dynamics of polymer solutions.
\newblock {\em Phys. Rev. Lett.}, 84:4765--4768.

\bibitem[Balkovsky et~al., 2001]{Balkovsky01}
Balkovsky, E., Fouxon, A., and Lebedev, V. (2001).
\newblock Turbulence of polymer solutions.
\newblock {\em Phys. Rev. E}, 64:056301.

\bibitem[Bird et~al., 1987]{Bird87}
Bird, R.~B., Armstrong, R.~C., and Hassager, O. (1987).
\newblock {\em Dynamics of polymeric liquids}, volume I. (Fluid mechanics), II.
  (Kinetic theory).
\newblock John Wiley \& Sons, New York.

\bibitem[Chertkov, 2000]{Chertkov00}
Chertkov, M. (2000).
\newblock Polymer stretching by turbulence.
\newblock {\em Phys. Rev. Lett.}, 84:4761.

\bibitem[Cranston and Scheutzow, 2002]{Cranston02}
Cranston, M. and Scheutzow, M. (2002).
\newblock Dispersion rates under finite mode kolmogorov flows.
\newblock {\em Ann. Appl. Prob.}, 12:511.

\bibitem[Dombre et~al., 1986]{Dombre86}
Dombre, T., Frisch, U., Greene, J.~M., Henon, M., Mehr, A., and Soward, A.~M.
  (1986).
\newblock Chaotic streamlines in the {ABC} flows.
\newblock {\em J. Fluid. Mech.}, 167:353--391.

\bibitem[Eckart, 1948]{Eckart48}
Eckart, C. (1948).
\newblock An analysis of the stirring and mixing processes in incompressible
  fluids.
\newblock {\em J. Mar. Res.}, 7:265.

\bibitem[Eckhardt, 2003]{Eckhardt03}
Eckhardt, B. (2003).
\newblock Echoes in classical dynamical systems.
\newblock {\em J. Phys. A}, 36:371.

\bibitem[Eckhardt and Hasco\"et, 2002]{Eckhardt02a}
Eckhardt, B. and Hasco\"et, E. (2002).
\newblock Chaotic advection by viscous dephasing.
\newblock {\em preprint}.

\bibitem[Eckhardt et~al., 2002]{Eckhardt02}
Eckhardt, B., Kronj{\"a}ger, J., and Schumacher, J. (2002).
\newblock Stretching of polymers in a turbulent environment.
\newblock {\em Comp. Phys. Commun.}, 147:538--543.

\bibitem[Falkovich et~al., 2002]{Falkovich02}
Falkovich, G., Gawedzki, K., and Vergassola, M. (2002).
\newblock Particles and fields in fluid turbulence.
\newblock {\em Rev. Mod. Phys.}, 73:913.

\bibitem[Fereday et~al., 2002]{Vassilicos02}
Fereday, D.~R., Haynes, P.~H., Wonhas, A., and Vassilicos, J.~C. (2002).
\newblock Scalar variance decay in chaotic advection and batchelor-regime
  turbulence.
\newblock {\em Phys. Rev. E}, 65:35301.

\bibitem[Homsy et~al., 2000]{Homsy00}
Homsy, G.~M., Aref, H., Breuer, K.~S., Hochgreb, S., Koseff, J.~R., Munson,
  B.~R., and Powell, K.~G. (2000).
\newblock Multi-media fluid mechanics.
\newblock {\em CD-ROM Cambridge University Press}.

\bibitem[J\"utner et~al., 1997]{Tabeling97}
J\"utner, B., Marteau, D., Tabeling, P., and Thess, A. (1997).
\newblock Numerical simulations of experiments on quasi-two-dimensional
  turbulence.
\newblock {\em Phys. Rev. E}, 55:5479.

\bibitem[Klages, 2002]{Klages02}
Klages, R. (2002).
\newblock Transitions from deterministic to stochastic diffusion.
\newblock {\em Europhys. Lett.}, 57:796.

\bibitem[Klages and Dorfman, 1995]{Klages95}
Klages, R. and Dorfman, J. (1995).
\newblock Simple maps with fractal diffusion coefficients.
\newblock {\em Phys. Rev. Lett.}, 74:387.

\bibitem[Klages and Dorfman, 1999]{Klages99}
Klages, R. and Dorfman, J. (1999).
\newblock Simple deterministic dynamical systems with fractal diffusion
  coefficients.
\newblock {\em Phys. Rev. E.}, 59:5361.

\bibitem[Klages and Dorfman, 1997]{Klages97}
Klages, R. and Dorfman, J.~R. (1997).
\newblock Dynamical crossover in deterministic diffusion.
\newblock {\em Phys. Rev. E.}, 55:1247.

\bibitem[Klyatskin et~al., 1996]{Klyatskin96}
Klyatskin, V.~I., Woyczynski, W.~A., and Gurarie, D. (1996).
\newblock Diffusing passive tracers in random imcompressible flows: statistical
  topography aspect.
\newblock {\em J. Stat. Phys.}, 84:797.

\bibitem[Kronj{\"a}ger et~al., 2002]{Kronjaeger02}
Kronj{\"a}ger, J., Braun, W., and Eckhardt, B. (2002).
\newblock Polymer stretching in stationary flows.
\newblock {\em preprint}.

\bibitem[Lumley, 1969]{Lumley69}
Lumley, J.~L. (1969).
\newblock Drag reduction by additives.
\newblock {\em Ann. Rev. Fluid Mech.}, 1:367.

\bibitem[Maxey and Riley, 1983]{Maxey83}
Maxey, M.~R. and Riley, J.~J. (1983).
\newblock Equation of motion for small rigid sphere in a nonuniform flow.
\newblock {\em Phys. Fluids}, 26:883.

\bibitem[Ottino, 1989]{Ottino89}
Ottino, J.~M. (1989).
\newblock {\em The kinematics of Mixing : Stretching, Chaos and Transport}.
\newblock Cambridge University Press.

\bibitem[Pierrehumbert, 1994]{Pierrehumbert94}
Pierrehumbert, R. (1994).
\newblock Tracer microstructure in the large-eddy dominated regime.
\newblock {\em Chaos, Solitons, Fractals}, 4:347.

\bibitem[Rothstein et~al., 1999]{Gollub99}
Rothstein, D., Henry, E., and Gollub, J.~P. (1999).
\newblock Persistent patterns in transient chaotic mixing.
\newblock {\em Nature}, 401:770--772.

\bibitem[Sornette and Cont, 1997]{Sornette97}
Sornette, D. and Cont, R. (1997).
\newblock Convergent multiplicative processes repelled from zero: Power laws
  and truncated power laws.
\newblock {\em J. Phys. I France}, 7:431--444.

\bibitem[Taylor, 1960]{Taylor60}
Taylor, G. (1960).
\newblock Low reynolds numbers flow.
\newblock {\em (16mm film).(Newton, Massachusetts: Educational Services
  Incorporated.)}.

\bibitem[Virk, 1975]{Virk75}
Virk, P.~S. (1975).
\newblock Drag reduction fundamentals.
\newblock {\em AIChE J.}, 21:625--656.

\bibitem[Williams et~al., 1997]{Gollub97}
Williams, B.~S., Marteau, D., and Gollub, J.~P. (1997).
\newblock Mixing of a passive scalar in magnetically forced two-dimensional
  turbulence.
\newblock {\em Phys. Fluids}, 9:2061--2080.

\end{thebibliography}
\chapbblname{paper}
\chapbibliography{paper.bib}
\nocite{*}

\end{document}